\documentclass[twoside,british]{iopart}
\usepackage[T1]{fontenc}
\usepackage[latin9]{inputenc}
\usepackage{color}
\usepackage{amstext}
\usepackage{amssymb}
\usepackage{graphicx}

\makeatletter
\usepackage{iopams}
\usepackage{setstack}

\usepackage{babel}
\begin{document}

\title[Pulling a DNA molecule through a nanopore embedded in an anionic membrane]{Pulling a DNA molecule through a nanopore embedded in an anionic membrane: tension propagation coupled to electrostatics}

\author{Jalal Sarabadani$^{1}$, Sahin Buyukdagli$^{2}$, and Tapio Ala-Nissila$^{3,4}$}

\address{$^1$School of Nano Science, Institute for Research in Fundamental Sciences (IPM), 19395-5531, Tehran, Iran}
\address{$^2$ Department of Physics, Bilkent University, Ankara 06800, Turkey}
\address{$^3$ Department of Applied Physics and QTF Center of Excellence, Aalto University, P.O. Box 11000, FI-00076 Aalto, Espoo, Finland}
\address{$^4$ Interdisciplinary Centre for Mathematical Modelling and Department of Mathematical Sciences, Loughborough University, Loughborough, Leicestershire LE11 3TU, United Kingdom}
\ead{jalal@ipm.ir}
\ead{buyukdagli@fen.bilkent.edu.tr}
\ead{Tapio.Ala-Nissila@aalto.fi}

\begin{abstract}
We consider the influence of electrostatic forces on driven translocation dynamics of a flexible polyelectrolyte
being pulled through a nanopore by an external force on the head monomer. To this end, we  
augment the iso-flux tension propagation (IFTP) theory with electrostatics for a negatively charged biopolymer pulled 
through a nanopore embedded in a similarly charged anionic membrane\footnote{A technical error in the interpretation of the sign of the charge distribution of the membrane has been fixed in the second and third updated versions.}. We show that in the realistic case of
a single-stranded DNA molecule, dilute salt conditions characterized by weak charge screening, and a negatively charged membrane, the translocation dynamics is unexpectedly accelerated despite the presence of large repulsive 
electrostatic interactions between the polymer coil on the {\it cis} side and the charged membrane.
This is due to the rapid release of the electrostatic potential energy of the coil during translocation, leading to an
effectively attractive force that assists end-driven translocation.
The speedup results in non-monotonic polymer length and membrane charge dependence of the exponent $\alpha$ characterizing 
the translocation time $\tau \propto N_0^\alpha$ of the polymer with length $N_0$. In
the regime of long polymers $N_0\gtrsim500$, the translocation exponent exceeds its upper limit $\alpha=2$ previously 
observed for the same system without electrostatic interactions.
\end{abstract}


\maketitle


\section{Introduction} \label{introduction}

Polymer translocation through a nanopore has been subject of numerous studies during the last two decades 
\cite{Muthukumar_book,Milchev_JPCM,Tapio_review,jalalJPC_2018,Review_Polymers_2019} following the 
seminal works by Bezrukov {\it et al.} in 1994 \cite{parsegian} and later by Kasianowicz {\it et al.} 
in 1996 \cite{kasi1996}. It has many technological applications in rapid DNA sequencing 
\cite{kasi1996,meller_2008,Luo_PRL_2008,Sigalov_Nano_Lett_2008,Ramin_PRX_2012}, 
drug delivery \cite{meller2003} and gene therapy. 
Motivated by these, many experimental 
\cite{branton_PRL_2003,branton2008,storm2005,meller_biophys_j_2004,wanunu_biophys_j_2014,Keyser_NatPhys_2006} 
as well as theoretical \cite{sung1996,muthu1999,chuang2001,metzler2003,grosberg2006,sakaue2007,rowghanian2011,%
dubbeldam2011,ikonen2012a,ikonen2012b,ikonen2013,jalal2014,jalal2015,jalal2017SR,jalal2017EPL,sakaue2016,%
Menais_SciRep_2016,unbiased_Slater_3,unbiased_Slater_2,Huopaniemi_2007,%
Research_Polymers_2018,Polymers_Oshanian_2018} 
works have been published in this research field during the last twenty years or so.
\begin{figure*}
    \begin{center}
        \includegraphics[width=1.0\textwidth]{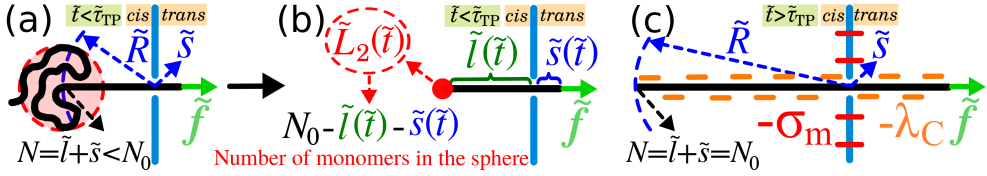}
    \end{center}
\caption{(a) A schematic of end-pulled polymer translocation through a nanopore in the tension propagation stage.
$\tilde{R}$ and $\tilde{s}$ are the distances of the tension front from the pore entrance and translocation coordinate, 
respectively. $N = \tilde{l} + \tilde{s} < N_0$ is the number of beads that have {been} already influenced by the tension, 
$\tilde{l}$ is the number of beads in the mobile subchain on the {\it cis} side, and $N_0$ in the polymer contour length.
The external driving force $\tilde{f}$ acts on the head monomer of the polymer only. 
(b) The electric charge inside the polymer coil (sphere) in panel (a) (highlighted 
by the red color) is approximated by a point charge (red dot) including 
$\tilde{L}_2 (\tilde{t})=N_0 - \tilde{l} (\tilde{t}) - \tilde{s} (\tilde{t})$ monomers. 
(c) The same as panel (a) but in the post propagation stage, where the tension has already reached the chain end
and $\tilde{l} + \tilde{s} = N_0$. 
$ {-} \sigma_{\textrm{m}} {<} 0$ and $-\lambda_{\textrm{C}} < 0$ are negative surface and 
linear charge densities distributed on the membrane and on the polymer, respectively 
(not shown in panels a and b).
}
\label{figure_1}
\end{figure*}

From a theoretical point of view, many studies have been devoted to elucidate the physics of 
uncharged polymer translocation through a nanopore embedded in an uncharged membrane when driven by
an external force.
In 2007 Sakaue came up with the idea of tension propagation (TP) in the context of driven polymer translocation 
\cite{sakaue2007}. A quantitative theory was developed starting 
in 2012 when Ikonen {\it et al}. showed that driven translocation processes can be described 
by using the TP formalism in the context of Brownian dynamics \cite{ikonen2012a,jalal2014}.
Combining the TP formalism and the iso-flux assumption (IFTP) \cite{rowghanian2011},
several different scenarios such as pore-driven 
translocation of a flexible and semi-flexible chain through a nanopore \cite{ikonen2012a,jalal2014,jalal2017SR}, 
pore-driven flexible polymer translocation 
under an alternating driving force through a flickering pore \cite{jalal2015}, and end-pulled polymer 
translocation through a nanopore \cite{jalal2017EPL}, have been investigated. 

In the TP formalism for driven translocation, a tension front propagates along the backbone of the subchain on the {\it cis} side. 
This is called TP stage, where the tension has not reached the chain end yet. Therefore, the subchain on the {\it cis} side 
is divided into two parts, namely a mobile part wherein the monomers move with net non-zero velocities towards the pore, 
and an equilibrium part where the monomers' average velocities in a narrow window of time are close to zero due to the 
random thermal fluctuations caused by the solvent. 
Finally, the tension reaches the chain end and the post propagation (PP) stage starts. This lasts until 
the whole mobile subchain on the {\it cis} side is sucked into the pore.

In the pore-driven case, as the relative dielectric constant of the solvent is typically high ($\approx 80$) 
with respect to that of the membrane ($\approx 2-4$), it is a good approximation to assume that an external force 
acts on the monomers located inside the nanopore only. However, in many cases of biological and experimental interest 
it is relevant to consider polyelectrolytes that are charged, such as DNA. This requires taking into account both 
the presence of counterions in the solution and the dielectric properties of the membrane through which the polyelectrolyte 
is translocating. Recently there has been an intense effort to model polyeletrolyte translocation including the details of 
the pore electrohydrodynamics and/or electrostatic polymer-membrane interactions, however, with the price of neglecting 
conformational polymer fluctuations~\cite{sahin_2018III,Research_Polymers_2018,Review_Polymers_2019}. 
This simplified modeling has allowed to characterize the electrohydrodynamic mechanism of experimentally observed DNA mobility reversal by charge inversion~\cite{ExpInv} and pressure-voltage traps~\cite{ExpPr}, and also enabled to identify strategies for faster polymer capture and slower translocation required for accurate biosequencing.

In this letter, we undertake the ambitious task to develop a unified theory of polymer translocation accounting for both tension propagation in a flexible chain and the electrostatic interactions between an anionic polymer and a {like-charged}
membrane when pulled through the pore by an external force on the head monomer. Within this electrostatically augmented IFTP formalism, we characterize the interplay between the electrostatic forces on the polyelectrolyte and the effect of the ubiquitous tension propagation mechanism.


\section{Model} \label{model}

It is a {challenge} to include both electrostatic interactions and dynamics of tension propagation in a 
flexible chain on the same footing. During translocation a flexible polymer samples complicated time-dependent 
configurations leading to highly varying electrostatic interactions. To make this tractable, we will consider 
here a simplified model where we take into account the average interaction at any given moment during end-pulled 
translocation. Our theoretical model of an anionic polymer translocating through an anionic
membrane nanopore is 
depicted in Fig. \ref{figure_1}. The polymer is composed of $N_0$ monomers and the external driving force only 
acts on the head monomer. Within the framework of the IFTP theory, the charge-free version of this model has been 
studied in detail in Ref. \cite{jalal2017EPL}. As it has been shown in Refs. \cite{jalalJPC_2018,jalal2014}, 
in the pore-driven case, three different regimes exist for the {\it cis} side: the trumpet (TR), stem-flower (SF) and 
strongly stretching (SS) regimes, corresponding to weak, moderate and strong external driving force, respectively. 
It should be also added that
in the limit of very weak driving force, Sakaue has presented additional scenarios \cite{sakaue2016}. 
For the end-pulled case, as both the {\it cis} and {\it trans} side subchains contribute to the translocation dynamics, 
complicated scenarios  of multiple regimes are involved in the theory \cite{jalal2017EPL}. 
Here, as the main goal is to illustrate how the electrostatic interactions affects 
the average polymer translocation dynamics, for both the {\it cis} and {\it trans} side subchains, we consider only 
the SS regime characterized by an external driving force satisfying the inequality $\tilde{f} \gtrsim N_0$ 
(see below for the definitions of dimensionless variables denoted by tilde).

Here we also assume that the linear self-avoiding flexible polymer chain is negatively charged such as a DNA molecule, 
with a linear charge density of $-\tilde{\lambda}_{\textrm{C}}<0$. Moreover, we assume that the 
membrane carries a negative 
charge with surface density ${-}\sigma_{\rm m}{<}0$, which is uniformly distributed on the membrane and constant during the translocation process. 
The negative membrane charge is a consequence of a low degree of protonation occurring in the typical high 
pH conditions of translocation experiments~\cite{sahin_2018III}.

In the SS regime during the TP stage, the fully straightened {\it cis} and {\it trans} portions of the mobile subchain have 
lengths $\tilde{l}$ and $\tilde{s}$, respectively.
Then, the immobile part on the {\it cis} side is coiled as highlighted in Fig. \ref{figure_1}(a) in light red 
color. 
The coil is assumed to be approximately spherical in shape and will 
be modelled as a negative point charge whose position is approximated to be at the end
of the tension propagation front  $\tilde{R}\tilde{(t)}$ (red dot in Fig. \ref{figure_1}(b)). 
As depicted in Fig. \ref{figure_1}(b), the number of monomers inside the sphere is 
$\tilde{L}_2 (\tilde{t}) = N_0 - \tilde{l} (\tilde{t}) - \tilde{s} (\tilde{t})$. 
The boundary between the mobile polymer portion and the inert coil on the {\it cis} side is the position of the tension front,
located at distance $\tilde{R}$ from the pore entrance. The SS regime is characterised by the equality $\tilde{R} = \tilde{l}$. 
As time passes and tension propagates along the backbone of the chain, the number of monomers inside 
the sphere decreases, and therefore its size shrinks. The TP 
stage ends when the tension reaches the chain end and 
the sphere disappears. In the subsequent PP 
stage, the whole chain in both {\it cis} and {\it trans} sides is mobile 
and fully straightened.

To mathematically formulate the model presented above, we will first express the relevant physical parameters in 
dimensionless units, denoted by a  tilde and defined
as $\tilde{Z} \equiv Z / Z_u$ \cite{ikonen2012a,jalal2014,jalal2015,jalal2017EPL,jalalJPC_2018}. 
The length, time, friction, force, velocity and monomer flux are written in units of $s_u \equiv \sigma$, 
$t_u \equiv \eta \sigma^2 / (k_{\rm B} T)$, $\Gamma_u \equiv \eta$, $f_u \equiv k_{\rm B} T/\sigma$, and 
$\varphi_u \equiv k_{\rm B} T/(\eta \sigma^2)$, respectively, where $\sigma$ is the segment length 
(or the size of each bead), $k_{\rm B}$ the Boltzmann constant, and $T$ and $\eta$ the solvent
temperature and friction per monomer, respectively. 
The linear polymer charge density 
is expressed in units of $\lambda_{{\textrm{C}},u} \equiv  e / \sigma$, with the electron charge 
$e = 1.6 \times 10^{-19} $C and the Kuhn length of a single DNA strand $\sigma = 1.5$ nm.
Lennard-Jones units are used for quantities without the tilde, such as the friction, time and force. 

In the SS regime, the entropic force due to fluctuations of the coiled subchain {on} the {\it cis} side and straightened subchains {on} the {\it cis} and {\it trans} sides is negligible as compared to the total driving force {on} the right hand side of Eq.~(\ref{BD_equation}). Therefore, {as already verified and confirmed by extensive MD simulations~\cite{jalal2017SR,jalal2017EPL}}, it is a very good approximation to consider the translocation process of an end-pulled case by the IFTP theory without {an} entropic force here.
Within our electrostatically augmented IFTP formalism, the equation of motion for the translocation coordinate 
$\tilde{s}$ corresponding to the number of beads on the {\it trans} side reads
\begin{equation}
\tilde{\Gamma} (\tilde{t}) \frac{d \tilde{s}}{ d \tilde{t}} =
\tilde{f} + \tilde{f}_{\textrm{pm}}^{\textrm{a}}
\equiv  \tilde{f}_{\textrm{tot}},
\label{BD_equation}
\end{equation}
where $\tilde{\Gamma} (\tilde{t})$ stands for the effective friction coefficient, 
$\tilde{f}$ is the external driving force acting on the head monomer of the polymer and oriented from the {\it cis} 
to {\it trans} side, $\tilde{f}_{\textrm{pm}}^{\textrm{a}}$
is the electrostatic force due to the interaction between the like
charges of the polymer and the membrane, the {superscript} $\textrm{a}=\{\rm{TP,PP}\}$ indicates the translocation stage, and $\tilde{f}_{\textrm{tot}}$ 
is the total force. The effective friction that contains the essential physics of the tension propagation theory is defined 
as $\tilde{\Gamma} (\tilde{t}) = \tilde{\eta}_{\rm{p}}+\tilde{\eta}_{\rm cis} (\tilde{t}) 
+ \tilde{\eta}_{\textrm{TS}} (\tilde{t})$, where $\tilde{\eta}_{\rm{p}}$ stands for the pore friction. 
Then, $\tilde{\eta}_{\rm cis} (\tilde{t}) $ and $ \tilde{\eta}_{\textrm{TS}} (\tilde{t})$ are the solvent friction coefficients 
associated with the mobile subchain on the {\it cis} and {\it trans} sides, respectively. In the SS regime where the {\it cis} 
and {\it trans} mobile portions of the polymer are straight lines, the hydrodynamic friction coefficient on each side is proportional 
to the length of the corresponding polymer portion, i.e.
$\tilde{\eta}_{\rm cis} (\tilde{t}) = \tilde{l} (\tilde{t}) = \tilde{R} (\tilde{t})$ and $\tilde{\eta}_{\rm TS} (\tilde{t}) = \tilde{s} (\tilde{t})$. 
Therefore, the total effective friction is given by
\begin{equation}
\tilde{\Gamma} (\tilde{t}) = \tilde{R} (\tilde{t}) + \tilde{s} (\tilde{t}) + \tilde{\eta}_{\rm{p}}.
\label{friction}
\end{equation}
\begin{figure*}[t]
    \begin{center}
        \includegraphics[width=0.95\textwidth]{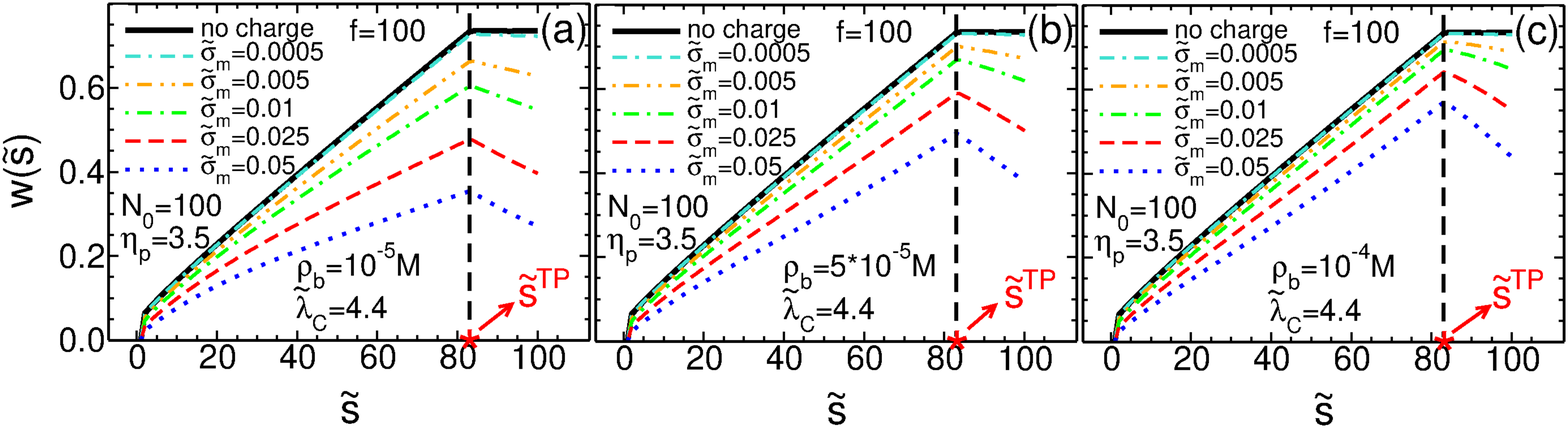}
    \end{center}
\caption{(a) The waiting time $w (\tilde{s})$ as a function of the translocation coordinate $\tilde{s}$ for various 
magnitudes of the negative
membrane charge density $\tilde{\sigma}_{\textrm{m}}$ with constant chain length 
$N_0 = 100$, pore friction $\eta_{\rm p} =3.5$, external driving force $f = 100$, salt concentration $\rho_{\textrm{b}} = 10^{-5}$ M, 
and the negative polymer line charge density 
$\lambda_{\textrm{C}} =  1/3.4~e/${\AA} ($\tilde{\lambda}_{\textrm{C}} =  4.4$) 
of a single-stranded DNA molecule~\cite{Research_Polymers_2018}. 
Panels (b) and (c) are the same 
as (a) but for different values of the salt concentrations $\rho_{\textrm{b}} = 5 \times 10^{-5}$ M, and $10^{-4}$ M, respectively.
In all panels the translocation coordinate at the TP time is denoted by $\tilde{s}^{\textrm{TP}}$ in red color.
}
\label{fig_WT}
\end{figure*}

In Eq. (\ref{BD_equation}), the electrostatic force on the polyelectrolyte reads 
$\tilde{f}_{\textrm{pm}}^{\textrm{a}} = - \partial \tilde{\Omega}_{\textrm{pm}}^{\textrm{a}} / \partial \tilde{s}$, 
where $\tilde{\Omega}_{\textrm{pm}}^{\textrm{a}}$ stands for the electrostatic polymer-membrane interaction energy. 
The latter is given in the TP and PP stages by~\cite{Research_Polymers_2018}, 
\begin{eqnarray}
\label{tp_energy}
\tilde{\Omega}_{\textrm{pm}}^{\textrm{TP}} &=& 
\frac{ 2 \tilde{\lambda}_{\textrm{C}} }{  \tilde{\mu} \tilde{\kappa}_{\textrm{b}}^2 } 
\big[ 1 - \textrm{e}^{- \tilde{\kappa}_{\textrm{b}} \tilde{l} (\tilde{t}) } 
\big]  
+ \frac{ 2 \tilde{\lambda}_{\textrm{C}} }{  \tilde{\mu} \tilde{\kappa}_{\textrm{b}}^2 } 
\big[ 1 - \textrm{e}^{- \tilde{\kappa}_{\textrm{b}} \tilde{s} (\tilde{t}) } \big]
-  \tilde{\lambda}_{\textrm{C}} L_2 (\tilde{t}) \tilde{\phi}_{\textrm{m}} [\tilde{l} (\tilde{t})];\nonumber\\
\tilde{\Omega}_{\textrm{pm}}^{\textrm{PP}}&=& 
\frac{ 2 \tilde{\lambda}_{\textrm{C}} }{  \tilde{\mu} \tilde{\kappa}_{\textrm{b}}^2 } 
\big[ 1 - \textrm{e}^{- \tilde{\kappa}_{\textrm{b}} \tilde{l} (\tilde{t}) } 
\big] 
+ \frac{ 2 \tilde{\lambda}_{\textrm{C}} }{  \tilde{\mu} \tilde{\kappa}_{\textrm{b}}^2 } 
\big[ 1 - \textrm{e}^{- \tilde{\kappa}_{\textrm{b}} \tilde{s} (\tilde{t}) } \big],
\label{mp_energy}
\end{eqnarray}
where the first and the second terms in $\tilde{\Omega}_{\textrm{pm}}^{\textrm{TP}}$ and 
$\tilde{\Omega}_{\textrm{pm}}^{\textrm{PP}}$ stand for mobile {\it cis} side subchain-membrane and mobile {\it trans} 
side subchain-membrane interactions including $\tilde{l}$ and $\tilde{s}$ beads, 
respectively, and the third term in $\tilde{\Omega}_{\textrm{pm}}^{\textrm{TP}}$ 
shows the sphere-membrane interaction that includes $\tilde{L}_2 (\tilde{t}) = N_0 - \tilde{l} -\tilde{s}$ beads.
Moreover, we used the Debye-H\"{u}ckel {(DH)} screening parameter $\kappa_{\textrm{b}} = 
\sqrt{ 8 \pi \ell_{\textrm{B}} \rho_{\rm b} }$ and the Gouy-Chapman length $\mu= e/(2 \pi \ell_{\textrm{B}}|\sigma_{\textrm{m}}|)$, 
with the Bjerrum length $\ell_{\textrm{B}} = e^2/(4\pi\epsilon_0\epsilon_{\textrm{w}}k_{\textrm{B}}T)\approx7$ 
{\AA} at solvent temperature  $T = 300$ K and permittivity $\epsilon_{\textrm{w}} = 80$, 
and the salt concentration $\rho_{\textrm{b}}$. In addition, $L_2 (\tilde{t}) = N_0 - \tilde{l} (\tilde{t}) - \tilde{s} (\tilde{t})$ 
is the number of beads in the coiled sphere, and 
$\tilde{\phi}_{\textrm{m}} [\tilde{l} (\tilde{t})] = -
2 \textrm{e}^{- \tilde{\kappa}_{\textrm{b}} \tilde{l} (\tilde{t})}/(\tilde{\mu}\tilde{\kappa}_{\textrm{b}})$ 
stands for the membrane potential derived within the DH theory, where $\tilde{l} (\tilde{t})$ {corresponds to} the distance of the sphere sub-chain to the nanopore, as the mobile sub-chain including $\tilde{l}$ beads in the {\it cis} side is fully straightened in the SS regime. Our DH approximation is motivated by 
the weak membrane charges considered in the present work. Although this approximation can be relaxed by introducing a 
charge renormalization procedure~\cite{Research_Polymers_2018}, in order to keep the analytical transparency of our model, 
we leave this improvement to a future work.
{Using the above equalities, the} contribution from the electrostatic force $\tilde{f}_{\textrm{pm}}^{\textrm{a}}$ 
to the force-balance equation~(\ref{BD_equation}) {follows} as
\begin{eqnarray}
\label{tp_force}
\tilde{f}_{\textrm{pm}}^{\textrm{TP}} &=& 
\frac{{-}2 \tilde{\lambda}_{\textrm{C}} }{ \tilde{\mu} \tilde{\kappa}_{\textrm{b}} }  
\bigg[ \textrm{e}^{- \tilde{\kappa}_{\textrm{b}} \tilde{s} (\tilde{t}) } 
- \hspace{-0.05cm}  \textrm{e}^{- \tilde{\kappa}_{\textrm{b}} \tilde{l} (\tilde{t}) } 
- \hspace{-0.05cm} \tilde{\kappa}_{\textrm{b}} L_2 (\tilde{t})   
\textrm{e}^{- \tilde{\kappa}_{\textrm{b}} \tilde{l} (\tilde{t}) }  
\frac{ \partial \tilde{l} (\tilde{t}) }{ \partial \tilde{s} (\tilde{t}) } \bigg],\nonumber\\
\tilde{f}_{\textrm{pm}}^{\textrm{PP}}  &=& 
\frac{{-}2 \tilde{\lambda}_{\textrm{C}} }{ \tilde{\mu} \tilde{\kappa}_{\textrm{b}} }  
\bigg[ \textrm{e}^{- \tilde{\kappa}_{\textrm{b}} \tilde{s} (\tilde{t}) }
- \textrm{e}^{- \tilde{\kappa}_{\textrm{b}} \tilde{l} (\tilde{t}) }
 \bigg].
\label{mp_force}
\end{eqnarray}
In the derivation of the second line in Eq.~(\ref{mp_force}), we used the mass conservation equation $N_0=l+s$ valid in the PP regime (see below) and implying the equality $\partial\tilde{l}/\partial \tilde{s}=-1$.
At this point, we note that a key novelty of our model is the non-trivial time-dependence of the derivative term 
$\partial \tilde{l} (\tilde{t})/\partial \tilde{s} (\tilde{t})$ in Eq.~(\ref{tp_force}) characterizing the TP regime.
The resulting function originating from the presence of the charged sphere accounts for the coupling between 
the electrostatic polymer-membrane interactions and the {non-uniform} tension propagation along the chain backbone. 
Thus, through this coupling, the present formalism goes beyond the purely stiff polymer model of 
Ref.~\cite{Research_Polymers_2018} where the TP regime is absent and the trivial equality 
$\partial\tilde{l}/\partial \tilde{s}=-1$ holds during the entire translocation process.

In order to evaluate the derivative $\partial \tilde{l} (\tilde{t})/\partial \tilde{s} (\tilde{t})$ in Eqs.~(\ref{tp_force}),
it should be noted that $\tilde{R}$ is equivalent to the end-to-end distance of the flexible self-avoiding
chain, i.e. $\tilde{R} = A_{\nu} N^{\nu}$, where $N = \tilde{l} + \tilde{s}$ is the number of all monomers that have 
been already influenced by the tension force (see Fig. \ref{figure_1}(a)), with the proportionality coefficient $A_{\nu} =1.15$
(for a coarse-grained bead-spring model as here), and the 3D Flory exponent $\nu=0.588$. 
Therefore, in the TP stage, the change in $\tilde{l} =\tilde{R}$ with respect to the translocation coordinate $\tilde{s}$ reads 
$\partial \tilde{R} / \partial \tilde{s}  = \nu A_{\nu}^{1/\nu} \tilde{R}^{(\nu -1 )/\nu} / 
\big[  1-\nu A_{\nu}^{1/\nu} \tilde{R}^{(\nu -1 )/\nu} \big] $.
At this point, we emphasize that in the TP stage, as the tension is propagating along the backbone of 
the chain located on the {\it cis} side, the length $\tilde{R}$ is growing, i.e. $\partial \tilde{R} / \partial \tilde{s} > 0$.
In contrast, in the PP stage where the straightened mobile subchain on the {\it cis} side is sucked into the pore, 
$\tilde{R}$ decreases and consequently one has $\partial \tilde{R} / \partial \tilde{s} = -1 < 0$.

Inserting the equalities above for $\partial \tilde{R} / \partial \tilde{s}$
together with the mass conservation in the TP and PP stages 
into Eqs.~(\ref{tp_force}), 
the IFTP Eqs.~(\ref{BD_equation}) and~(\ref{friction}) in the TP and PP stages can be expressed solely 
in terms of the coordinate $\tilde{l}$. The explicit form of the corresponding equations and asymptotic 
analytical predictions will be presented in a future work. 
Finally, the function $\tilde{l}(t)$ obtained from the numerical solution of Eqs.~(\ref{BD_equation}) and~(\ref{friction}) 
should be used in the scaling law $\tilde{l}=A_\nu(\tilde{l}+\tilde{s})^\nu$ and $N_0=\tilde{l}+\tilde{s}$ 
to obtain the time dependence of the translocation coordinate $\tilde{s}(t)$ 
in the TP and PP stages, respectively.


\section{Results}
\label{results}
In order to describe the dynamics of the translocation process at the monomer level, we first examine  the waiting time (WT) defined as
the time that each bead spends in the pore during translocation. We then illustrate 
the global dynamics of the translocation process by focusing on the translocation exponent $\alpha$, 
which is defined as $\tau \propto N_0^{\alpha}$, where $\tau$ is the average translocation time and $N_0$ the
contour length of the polymer.


\subsection{Waiting time} \label{WT}
To characterize the effect of polymer-membrane interactions on the translocation dynamics, the WT is plotted in Figs.~\ref{fig_WT}(a)-(c)
as a function of the translocation coordinate $\tilde{s}$. 
The plots also display in red the translocation coordinate $\tilde{s}^{\textrm{TP}}$ at the TP time corresponding to 
the time that takes for the tension front to reach the chain end. The WT curves {are} 
plotted for three different bulk salt concentrations 
ranging from $\rho_{\rm b}=10^{-5}$ M to $10^{-4}$ M, and various negative 
membrane charge density values with increasing magnitude of
$\sigma_{\rm m}$ from top to bottom (see the legends). We consider here
the dilute salt regime where the electrostatic interactions are weakly screened and expected to play an important role. 
It should be noted that this low salt concentration regime has been previously reached by translocation experiments~\cite{ExpInv}.
The line charge density of the polymer is set to the value $\lambda_{\textrm{C}} =  1/3.4~{\rm e}/${\AA} 
corresponding to a single-stranded DNA molecule. The remaining model parameters are given in the caption of Fig.~\ref{fig_WT}. 
We finally emphasize that because the electrostatic force in Eq.~(\ref{tp_force}) 
involves simply 
the product of the membrane and polymer charge densities, the results of our manuscript can be extrapolated to polyelectrolytes 
of different charge strengths by rescaling the membrane charge density value.
\begin{figure*}[t]
    \begin{center}
        \includegraphics[width=1.0\textwidth]{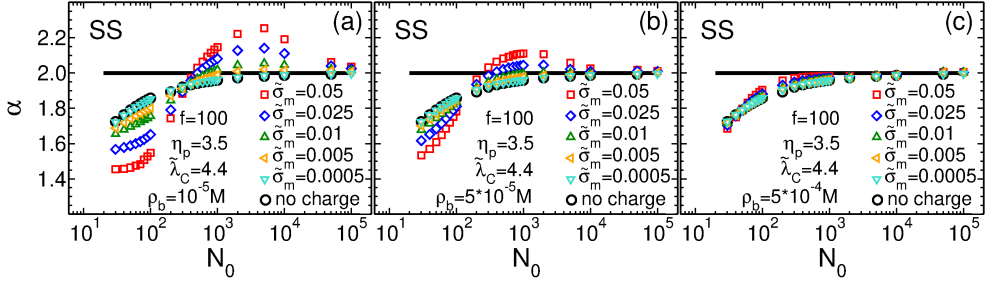}
    \end{center}
\caption{(a) Translocation time exponent $\alpha$ as a function of the polymer contour length for various values of 
the membrane charge density. The black horizontal line is the asymptotic limit $\alpha=2$ for {the} uncharged {system}.
The model parameters given in the legend are the same as in Fig. \ref{figure_1}.
}
\label{fig_exponent}
\end{figure*}

Figures~\ref{fig_WT}(a)-(c) show that at constant salt concentration, increasing the magnitude of the membrane charge density {\it decreases} 
the WT, i.e. $\sigma_{\rm m}\uparrow w(s)\downarrow$. 
Consequently, the TP time corresponding to the integral of the WT curve from zero to $\tilde{s}^{\textrm{TP}}$ decreases 
with the increase of the membrane charge strength. This is a surprising result as one would naively expect the electrostatic 
repulsion between the chain and the membrane to slow down rather than speed up the translocation speed.
The physical mechanism behind this seemingly counterintuitive effect will be discussed below. The comparison of Figs.~\ref{fig_WT}(a)-(c) also indicates that at constant membrane charge density, 
added bulk salt rapidly screens out the electrostatic polymer-membrane {repulsion}, as expected. One indeed notes that the increment of the salt concentration leads to an increase of the WT and reduction of the translocation rate, i.e. $\rho_{\rm b}\uparrow w(s)\uparrow$. 
Interestingly, the translocation coordinate $\tilde{s}^{\textrm{TP}}$ at the TP time remains unaffected 
by the value of the surface charge density and the salt concentration. This indicates that the electrostatics {does} 
not directly 
affect where the initiation of the post-propagation occurs in the chain.


\subsection{Translocation time exponent} \label{exponent}
Accurate polymer sequencing by translocation requires the extension of the ionic current blockade caused by the 
translocating polymer and thus the prolongation of the translocation event. Therefore, it is essential to characterize 
the effect of the experimentally tunable electrostatic model parameters on the total translocation time $\tau$ for 
the entire polymer chain to translate from the {\it cis} to the {\it trans} side.
 
As seen in Figs. \ref{fig_WT}(a)-(c), at constant salt concentration, the translocation time corresponding to the integral 
of the WT time curve over the translocation coordinate $s$ drops with the increase of the magnitude of the membrane charge, i.e. 
$\sigma_{\rm m}\uparrow \tau\downarrow$. Moreover, the comparison of the panels (a)-(c) at constant surface charge 
indicates that due to electrostatic screening, the addition of bulk salt increases the translocation time and slows 
down the translocation dynamics, i.e. $\rho_{\rm b}\uparrow \tau\uparrow$.

\begin{figure*}[t]
    \begin{center}
        \includegraphics[width=0.85\textwidth]{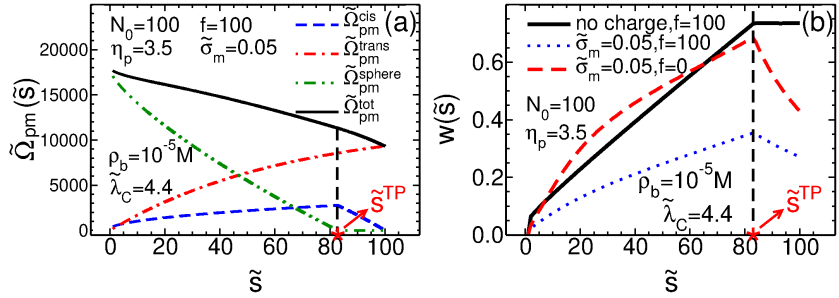}
    \end{center}
\caption{
(a) The electrostatic interaction energy $\tilde{\Omega}_{\textrm{pm}}$ and its components (denoted by the corresponding superscripts) 
between the polymer and the like-charged membrane as a function of the translocation coordinate $\tilde{s}$, with the same 
values of the parameters as in panel (a) in Fig.~\ref{fig_WT}, but for a fixed value of the surface charge density 
$\tilde{\sigma}_{\textrm{m}} = 0.05$. The blue dashed and the red dash-dotted 
lines display the contributions from the mobile {\it cis} and {\it trans} sides subchains, 
respectively, while the contribution from the sphere is shown by the green dash-dotted-dotted line.
The total electrostatic energy is repulsive but monotonically decreases (black solid line).
This stems from a rapid decrease of the sphere contribution.
(b) The WT as a function of $\tilde{s}$ for the uncharged system with an external driving force $f=100$ (black solid line) and for 
a charged system ($\tilde{\sigma}_{\textrm{m}} = 0.05$) with external driving force $f=100$ (blue dotted line) 
and $f=0$ (red dashed line), respectively. See text for details.
}
\label{fig_energy}
\end{figure*}

An exact scaling form for the translocation time in the SS regime for an end-pulled polymer chain has been derived in Ref. 
\cite{jalal2017EPL}. In the asymptotic limit of very long chains, the translocation time exponent 
defined as $\tau \propto N_0^\alpha$ is $\alpha = 2$, with large correction-to-scaling terms 
arising from the pore friction and the {\it trans} side friction of the chain. In order to understand the effect of 
electrostatic polymer-membrane interactions on the dependence of the translocation time on the contour length, 
we illustrate  in Fig. \ref{fig_exponent} the effective translocation time exponent $\alpha$ as a function of $N_0$ 
for various membrane charge and salt strengths.
In agreement with the results above, the comparison of Figs. \ref{fig_exponent} (a)-(c) shows that as the increase of 
the salt concentration suppresses electrostatic interactions in the system, the trend of the translocation exponent 
for different values of the surface charge density becomes similar to that of an uncharged system (black open circles). 
As a result, the non-trivial behaviour of the exponent $\alpha$ originating from electrostatic interactions appears in the highly dilute salt concentration 
regime $\rho_{\rm b}\lesssim10^{-4}$ M. 

Figures \ref{fig_exponent}(a) and (b) show that due to the same electrostatic polymer-membrane interactions, as the surface 
charge density increases, the translocation exponent becomes a non-monotonic function of the membrane charge 
$\sigma_{\rm m}$ and the chain length $N_0$. More precisely, for short polymers with length $N_0\lesssim500$, the membrane charge strength
{decreases} the scaling exponent, i.e. $\sigma_{\rm m}\uparrow\alpha\downarrow$. For long polymers $N_0\gtrsim500$, the exponent 
increases with the membrane charge ($\sigma_{\rm m}\uparrow\alpha\uparrow$) and exceeds its upper bound $\alpha=2$ observed 
for the uncharged system. We now focus on the non-monotonic polymer length dependence of the exponent $\alpha$. One sees that 
with the increase of the polymer length, the exponent initially rises ($N_0\uparrow\alpha\uparrow$), reaches a peak at the 
characteristic length $N_0^*$, and subsequently drops in the long polymer regime ($N_0\uparrow\alpha\downarrow$). We finally
note that the characteristic length for the maximum of the exponent $\alpha$ is lowered by added salt, i.e. 
$\rho_{\rm b}\uparrow N_0^*\downarrow$.

\subsection{Physical mechanism behind repulsive speedup} \label{exponent}

To understand the physical mechanism behind the unexpected speedup of the translocation process with repulsive interactions
we need to carefully consider the various contributions to the total electrostatic energy (EE) in the system. In Fig.~\ref{fig_energy}(a) the 
total EE
$\tilde{\Omega}_{\textrm{pm}}$ and its various components have been plotted as a function of the translocation 
coordinate $\tilde{s}$ for fixed value of the membrane surface charge density $\tilde{\sigma}_{\textrm{m}} = 0.05$. 
As can be seen the total EE is positive for all values of the translocation coordinate $\tilde{s}$ as expected
(black solid line), but it unexpectedly monotonically {\it decreases} as $\tilde{s}$ increases. This results in a total effective force that
is directed towards the {\it trans} side and assists end-driven translocation. 

To explicitly show the effect of the effective attractive force in Fig.~\ref{fig_energy}(b) we plot
the WT as a function of $\tilde{s}$
for an uncharged system (black solid line) and for two systems with 
surface charge density $\tilde{\sigma}_{\textrm{m}} = 0.05$. The blue dotted and the red dashed lines display 
WT for the charged systems with the values of the external driving force 
$f=100$ and $f=0$, respectively. Quite interestingly, even without an external driving force, i.e. $f=0$,
the translocation process is successful and also close to the case where $f=100$ for the uncharged system.

The physical reason behind the speedup can be seen from the EE components plotted
in Fig.~\ref{fig_energy}(a). The main contribution to the total repulsive EE comes from the charged sphere that comprises
most of the monomers at $t=0$. When the tension front starts propagating, the EE components from
the {\it cis} and {\it trans} sides increase at the expense of the charged sphere, whose energy rapidly drops with the
decreasing number of monomers in it. 
It is also important to point out that the unfolding of the coil modeled here as the melting of the corresponding 
charged sphere is not the its only degree of freedom; the sphere can also translate away from the membrane. Figure \ref{fig_energy}(a) shows that 
indeed it both melts and translates. The {\it cis} EE increases (blue curve), which means that the {\it cis} portion gets longer. This is possible only if the sphere 
located at the tension front moves away from the membrane, which would correspond to {its} repulsive interaction with the membrane wall ({blue} curve minimum at $\tilde{s}=0$).  However, the melting of the sphere brings a dominant attractive contribution (strongly decreasing green curve with minimum at $\tilde{s} = \tilde{s}_{\textrm{TP}}$). Consequently, the sphere's energy overall favors 
translocation towards the pore. This results in an overall reduction of the EE and an \textit{effective} attractive force that speeds up the translocation dynamics. 


Finally, in the PP stage the EE is governed only by the straightened {\it cis} and {\it trans} side subchains.
As the straightened {\it cis} side subchain is sucked into the pore and its length rapidly decreases,
the repulsive EE due to this part drops faster than that of the {\it trans} side subchain rises. This still leads to an overall
attractive force that assists translocation (see also Fig.~\ref{fig_energy}(b)).


\section{Summary and discussion} \label{summary}
An electrostatically augmented IFTP theory has been introduced to investigate end-pulled polyelectrolyte translocation 
through a nanopore embedded in a charged membrane. In this work, we have considered the case where there is a repulsive electrostatic
interaction between the polyelectrolyte and the charged membrane. The unexpected finding is that despite strong repulsive
interactions at low salt, the total EE of the polymer {\it decreases} as the tension front propagates in the chain, leading to an
effectively attractive force that actually speeds up the translocation dynamics. This is due to the fact that most of the EE is stored in
the coiled part of the chain located at the end of the tension front, and the EE of this charged sphere decreases much faster that the EE 
contributions from the {\it cis} and {\it trans} parts of the chain increase. The speedup leads to 
non-monotonic behavior of the translocation time 
exponent $\alpha$. For short polymers $N_0\lesssim500$, the increment of the membrane charge {strength} reduces the exponent 
$\alpha$ towards its lower bound $\alpha=1$. In the long polymer regime $N_0\gtrsim500$, the exponent increases with the 
membrane charge {strength} and exceeds its upper limit $\alpha=2$ previously observed for the charge-free system~\cite{jalal2017EPL}. 

We would like to highlight the main differences between the present formalism and previous electrostatic models of polymer translocation. Via the inclusion of the charged coil, the translocation model introduced herein is a direct extension of the purely electrostatic model of Ref.~\cite{Research_Polymers_2018}. In the latter model where the coil part had been neglected, the translocation process was assumed to be always in the PP regime. Thus, this important extension allowed us to consider here for the first time the electrostatics of the TP regime governed by the coil dynamics. We would like to emphasize as well the main difference between the present model and the electrohydrodynamic formalism of Ref.~\cite{sahin_2017}. It should be noted that the latter model was developed for the translocation of short polymer sequences where one can assume a stiff polymer configuration and thus bypass the presence of the coil portion. However, considering the comparable length of the polymer and the nanopore, the model of Ref.~\cite{sahin_2017} took explicitly into account the electrohydrodynamics of the translocation process as well as the electrostatic interactions between the polymer and the pore wall. In the present article where we focus on the translocation of polymers much longer than the nanopore, the electrohydrodynamic forces of relatively short range have been smeared out and included in terms of the pore friction coefficient $\tilde{\eta}_{\rm p}$ of Eq.~(\ref{BD_equation}). Future works can consider the inclusion of the coil dynamics and pore electrohydrodynamics on the same footing, but this challenging task is beyond the scope of the present article.

The results of our newly introduced model can be easily tested by current translocation experiments involving atomic force microscope (AFM)~\cite{Ritort_AFM} as well as magnetic or optical tweezers~\cite{Smith_Nature,Bulushev_1,Bulushev_2,Bulushev_3,Keyser_Nature_1,Keyser_Nature_2,Sischka}. Our predictions may also contribute to the improvement of our control over the polymer translocation dynamics in next generation biosensing techniques.

Finally, we would like to comment on the key assumptions made in the model. The main approximation here has been to assume that the
part of the chain unaffected by the tension front on the {\it cis} side can be replaced by a point charge whose position is 
always given by $\tilde{R}(\tilde{t})$ in the TP regime. Due to the high repulsive force stemming from the sphere-membrane interaction,
it is possible that its center of mass does not exactly follow $\tilde{R}(\tilde{t})$
but acquires an additional drift velocity component that moves it away from the membrane. 
If this were the case, the EE component from the sphere would decrease somewhat faster than our model predicts. 
Thus, it is possible that our theory slightly overestimates the total EE and the speedup. 
We will investigate this possibility
in future work.



\ack{}{}
T.A-N. has been supported in part by the Academy of Finland through its 
PolyDyna (no. 307806) and QFT Center of Excellence Program grants (no. 312298). We acknowledge the computational 
resources provided by the Aalto Science-IT project and the CSC IT Center for Science, Finland.


\section*{References}{}


\begin{thebibliography}{10}

\bibitem{Muthukumar_book} Muthukumar M 2011 {\it Polymer Translocation} (Taylor and Francis)

\bibitem{Milchev_JPCM} Milchev A 2011 {\it J. Phys.: Condens. Matter} {\bf 23} 103101 

\bibitem{Tapio_review} Palyulin V V, Ala-Nissila T and Metzler R 2014 {\it Soft Matter} {\bf 10} 9016

\bibitem{jalalJPC_2018} Sarabadani J and Ala-Nissila T 2018 {\it J. Phys.: Condens. Matter} {\bf 30} 274002

\bibitem{Review_Polymers_2019} Buyukdagli S, Sarabadani J and Ala-Nissila T 2019 {\it Polymers} {\bf 11} 118 

\bibitem{parsegian} Bezrukov S M, Vodyanoy I and Parsegian A V 1994 {\it Nature} {\bf 370} 279
  
\bibitem{kasi1996} Kasianowicz J J {\it et al.} 1996 {\it Proc. Natl. Acad. Sci.} {\bf 93} 13770
  
\bibitem{meller_2008} Meller A, Nivon L and Branton D. 2001 {\it Phys. Rev. Lett.} {\bf 86} 3435

\bibitem{Luo_PRL_2008} Luo K, Ala-Nissila T, Ying S -C and Bhattacharya A 2008 {\it Phys. Rev. Lett.} {\bf 100} 058101

\bibitem{Sigalov_Nano_Lett_2008} Sigalov G, Comer J, Timp G and Aksimentiev A 2008 {\it Nano Lett.} {\bf 8} 56
  
\bibitem{Ramin_PRX_2012} Cohen J A, Chaudhuri A and Golestanian R 2012 {\it Phys. Rev. X} {\bf 2} 021002
  
\bibitem{meller2003} Meller A 2003 {\it J. Phys. Condens. Matter} {\bf 15} R581

\bibitem{branton_PRL_2003} Sauer-Budge A F {\it et al.} 2003 {\it Phys. Rev. Lett.} {\bf 90} 238101

\bibitem{branton2008} Branton D {\it et al.} 2008 {\it Nature Biotech.} {\bf 26} 1146

\bibitem{storm2005} Storm A J {\it et al.} 2005 {\it Nano Lett.} {\bf 5} 1193
  
\bibitem{meller_biophys_j_2004} Mathe J {\it et al.} 2004 {\it Biophys. J.} {\bf 87} 3205

\bibitem{wanunu_biophys_j_2014} Carson S, Wilson J, Aksimentiev A and Wanunu M 2014 {\it Biophys. J.} {\bf 107} 2381

\bibitem{Keyser_NatPhys_2006} Keyser U F {\it et al.} 2006 {\it Nature Physics} {\bf 2} 473

\bibitem{sung1996} Sung W and Park P J 1996 {\it Phys. Rev. Lett.} {\bf 77} 783

\bibitem{muthu1999} Muthukumar M 1999 {\it J. Chem. Phys.} {\bf 111} 10371

\bibitem{chuang2001} Chuang J, Kantor Y and Kardar M 2001 {\it Phys. Rev. E} {\bf 65} 011802

\bibitem{metzler2003} Metzler R and Klafter J 2003 {\it Biophys. J.} {\bf 85} 2776

\bibitem{grosberg2006} Grosberg A Y {\it et al.} 2006 {\it Phys. Rev. Lett.} {\bf 96} 228105

\bibitem{sakaue2007} Sakaue T 2007 {\it Phys. Rev. E} {\bf 76} 021803

\bibitem{rowghanian2011} Rowghanian P and Grosberg A Y 2011 {\it J. Phys. Chem. B} {\bf 115} 14127
 
\bibitem{dubbeldam2011} Dubbeldam J L A {\it et al.} 2012 {\it Phys. Rev. E} {\bf 85} 041801

\bibitem{ikonen2012a} Ikonen T {\it et al.} 2012 {\it Phys. Rev. E}  {\bf 85} 051803

\bibitem{ikonen2012b} Ikonen T {\it et al.} 2012 {\it J. Chem. Phys.} {\bf 137} 085101

\bibitem{ikonen2013} Ikonen T {\it et al.} 2013 {\it Europhys. Lett} {\bf 103} 38001

\bibitem{jalal2014} Sarabadani J, Ikonen T and Ala-Nissila T 2014 {\it J. Chem. Phys.} {\bf 141} 214907

\bibitem{jalal2015} Sarabadani J, Ikonen T and Ala-Nissila T 2015 {\it J. Chem. Phys.} {\bf 143} 074905
  
\bibitem{jalal2017SR} Sarabadani J {\it et al.} 2017 {\it Sci. Rep.} {\bf 7} 7423

\bibitem{jalal2017EPL} Sarabadani J {\it et al.} 2017 {\it Europhys. Lett.} {\bf 120} 38004

\bibitem{sakaue2016} Sakaue T 2016 {\it Polymers} {8} 424

\bibitem{Menais_SciRep_2016} Menais T, Mossa S and Buhot A 2016 {\it Sci. Rep.} {\bf 6} 38558

\bibitem{unbiased_Slater_2} de Haan H W and Slater G W 2012 {\it J. Chem. Phys.} {\bf 136} 204902
  
\bibitem{unbiased_Slater_3} Gauthier M G and Slater G W 2009 {\it Phys. Rev. E} {\bf 79} 021802

\bibitem{Huopaniemi_2007} Huopaniemi I {\it et al.} 2007 {\it Phys. Rev. E} {\bf 75} 061912

\bibitem{Research_Polymers_2018} Buyukdagli S, Sarabadani J and Ala-Nissila T 2018 {\it Polymers} {\bf 10} 1242

\bibitem{Polymers_Oshanian_2018} Malgaretti P and Oshanin G 2019 {\it Polymers} {\bf 11(2)} 251

\bibitem{sahin_2018III} Buyukdagli S and Ala-Nissila T 2018 {\it Europhys. Lett.} {\bf 123} 38003

\bibitem{ExpInv} Qiu S {\it et al.} 2015 {\it Soft Matter} {\bf 11} 4099

\bibitem{ExpPr} Hoogerheide D P, Lu B and Golovchenko J A 2014 {\it ACS Nano} {\bf 8} 738

\bibitem{sahin_2017} Buyukdagli S and Ala-Nissila T 2017 {\it J. Chem. Phys.} {\bf 147} 114904

\bibitem{Ritort_AFM} Ritort F 2006 {\it J. Phys.: Condens. Matter} {\bf 18} R531

\bibitem{Smith_Nature} Smith D E {\it et al.} 2001 {\it Nature} {\bf 413} 748

\bibitem{Bulushev_1} Bulushev R D {\it et al.} 2014 {Nano Lett.} {\bf 14} 6606

\bibitem{Bulushev_2} Bulushev R D, Marion S and Radenovic A {\it et al.} 2015 {Nano Lett.} {\bf 15} 7118

\bibitem{Bulushev_3} Bulushev R D {\it et al.} 2016 {Nano Lett.} {\bf 16} 7882

\bibitem{Keyser_Nature_1} Keyser U F {\it el al.} 2006 {\it Nat. Phys.} {\bf 2} 473

\bibitem{Keyser_Nature_2} Keyser U F, Dekker N H, Dekker C and Lemay S G 2009 {\it Nat. Phys.} {\bf 5} 347

\bibitem{Sischka} Ritort F 2006 {\it J. Phys.: Condens. Matter} {\bf 18} R531


\end{thebibliography}
\end{document}